\documentclass[12pt,preprint]{aastex}
\usepackage{amsmath, amsthm, amssymb}

\newcommand{\Rs}{\ensuremath{R_{\odot}}}
\newcommand{\Ms}{\ensuremath{M_{\odot}}}
\newcommand{\beq}{\begin{equation}}
\newcommand{\eeq}{\end{equation}}
\newcommand{\Dbar}{\ensuremath{\overline D}}
\newcommand{\Rbar}{\ensuremath{\overline R}}
\newcommand{\fbar}{\ensuremath{\overline f}}
\newcommand{\nbar}{\ensuremath{\overline n}}
\newcommand{\zat}{\ensuremath{ {\zeta_a^t} }}
\newcommand{\Wan}{\ensuremath{{{\mathcal W}_{j}^N}}}
\newcommand{\Whan}{\ensuremath{{{\mathcal W}_{j+1/2}^N}}}
\newcommand{\Ncal}{\ensuremath{{\mathcal N}}}
\newcommand{\Dcal}{\ensuremath{{\mathcal D}}}
\newcommand{\eg}{{\it e.g.}}
\newcommand{\ie}{{\it i.e.}}
\newcommand{\viz}{{\it viz.}}

\slugcomment{To appear in Ap. J.}

\shorttitle{Binary Evolution in Globular Clusters - 
Stochastic Approach}
\shortauthors{Banerjee and Ghosh}

\begin{document}


\title{Evolution of Compact-Binary Populations\\
in Globular Clusters: A Boltzmann Study\\
II. Introducing Stochasticity}


\author{Sambaran Banerjee and Pranab Ghosh}
\affil{Department of Astronomy \& Astrophysics \\ 
Tata Institute of Fundamental Research, Mumbai 400 005, India}


\begin{abstract}
We continue exploration of the Boltzmann scheme started in Banerjee
and Ghosh (2007, henceforth Paper I) for studying the evolution of
compact-binary populations of globular clusters, introducing in
this paper our method of handling the stochasticity inherent in
dynamical processes of binary formation, destruction and hardening
in globular clusters. We describe these stochastic processes as
\emph{Wiener processes}, whereupon the Boltzmann equation becomes
a stochastic partial differential equation, the solution of which
requires the use of \emph{It\^{o} calculus} (this use being the
first, to our knowledge, in this subject), in addition to ordinary
calculus. We focus on the evolution of (a) the number of X-ray
binaries $N_{XB}$ in globular clusters, and (b) the orbital-period
distribution of these binaries. We show that, although the details
of the fluctuations in the above quantitities differ from one
``realization'' to another of the stochastic processes, the general
trends follow those found in the continuous-limit study of Paper I,
and the average result over many such realizations is close to the
continuous-limit result. We investigate the dependence of $N_{XB}$
found by these calculations on two essential globular-cluster
parameters, namely, the star-star and star-binary encounter-rate
parameters $\Gamma$ and $\gamma$, for which we had coined the
name Verbunt parameters in Paper I. We compare our computed results
with those from CHANDRA observations of Galactic globular clusters,
showing that the expected scalings of $N_{XB}$ with the Verbunt
parameters are in good agreement with the observed ones. We indicate
what additional features can be incorporated into the scheme in
future, and how more elaborate problems can be tackled.
\end{abstract}





\section{Introduction}
\label{intro}

In this series of papers, we are studying the evolution of compact-binary 
populations of globular clusters with the aid of a Boltzmann scheme 
which we introduced in \citet{pI}, henceforth Paper I. This scheme
follows compact-binary evolution as a result of both (a) those processes 
which determine compact-binary evolution in isolation (\ie, outside 
globular clusters), \eg, angular momentum loss by gravitational radiation 
and magnetic braking, as also orbital evolution due to mass transfer, 
and (b) those processes which arise from encounters of compact binaries 
with the dense stellar background in globular clusters, \eg, collisional 
hardening \citep{h75,sh79,bg2006}, binary formation through tidal capture
and exchange processes, and binary destruction \citep{fpr75,pt77,lo86,
dr92,dr94,spz,hb83}. We treat all of the above 
processes simultaneously through our Boltzmann scheme, the aim being to 
see their combined effect on the compact-binary population as a whole, in 
particular on the evolution of (a) the total number of X-ray binaries as 
the formation and destruction processes continue to operate, and, (b) the 
orbital-period distribution of the population. As stressed in Paper I, 
our scheme is the original Boltzmann one (not the Fokker-Planck reduction
of it), which, by definition, is capable of handling both the 
\emph{combined} small effects of a large number of frequent, weak, distant 
encounters and the \emph{individual} large effects of a small 
number of rare, strong, close encounters \emph{on the same footing}.
We note here that, although Monte Carlo Fokker-Planck approaches were 
normally thought to be capable of handling only the former effects, 
schemes for including the latter have been proposed and studied 
recently \citep{fr2003,fr2007}.     

In Paper I, we studied the problem in the continuous limit, wherein we  
used continuous representations for both kinds of processes described
above, \ie, those of category (a) above, which are inherently continuous, 
and also those of category (b), which are inherently stochastic. For the   
latter category, therefore, we used the \emph{continuous limit} of the 
above stochastic processes, wherein the probability or cross-section of 
a particular such process happening with a given set of input and output 
variables was treated as a continuous function of these variables. These
cross-sections were, of course, those that had been determined from 
extensive numerical experiments with two-body and three-body encounters 
performed earlier \citep{hhm96,z297}. 

In this paper, we address the next question, namely, how is the inherent  
stochasticity of the processes of category (b) to be introduced into our
scheme, to be handled simultaneously with the inherently continuous 
nature of those of category (a)? As stressed in Paper I, this step is of
great importance, since it is a simultaneous operation of the above 
continuous and stochastic processes in globular clusters that leads to the 
observed properties of compact-binary populations in them. To this end,
we introduce stochasticity into our Boltzmann study in this paper in the
following way. For a first look, we consider the rates of the processes 
of category (b) as \emph{randomly fluctuating} about the mean rates 
described in Paper I, while those of the processes of category (a) 
remain continuous, as before. We model these fluctuations as a 
\emph{Wiener process} (see Appendix A and references therein), which is 
the mathematical description of Brownian motion. 

With this prescription, the Boltzmann equation governing the evolution 
of the distribution function $n(a,t)$ of compact binaries in time $t$ 
and orbital radius $a$ becomes a \emph{stochastic} partial differential 
equation (henceforth SPDE), instead of the ordinary partial differential 
equation (henceforth OPDE) which it was in the continuous limit. We 
handle the solution of this SPDE with the aid of techniques developed 
largely during the last fifteen years \citep{kl94,g95,ok2004}. These 
techniques involve the use of the \emph{It\^{o} calculus} (see Appendix 
B and references therein), instead of ordinary calculus, for handling 
the stochastic terms.  

Our results show that the full solutions with stochasticity included
have fluctuations which vary from one ``realization'' to another of
the stochastic processes, as expected. However, the full results show
trends which generally follow those in the continuous limit. 
Furthermore, the average result over many realizations comes very 
close to the continuous limit, showing the importance of the latter
limit for understanding mean trends. On the other hand, understanding
fluctuations in a typical full run is also very important, as this
gives us a first idea of the magnitude of fluctuations we can expect 
in the data on X-ray binaries in globular clusters as a result of the
stochastic processes, as also the expected trends in the fluctuations 
with the essential globular-cluster parameters, \eg, the Verbunt
parameters introduced in Paper I (also see below).     

Comparison of our computed trends in the number $N_{XB}$ of X-ray
binaries in Galactic globular clusters with the Verbunt parameters
on the one hand, with observed trends in recent CHANDRA data on 
Galactic globular clusters on the other, shows that our full results 
are in good agreement with observation. We have thus constructed a 
straightforward, very inexpensive scheme for following the evolution 
of compact-binary populations in globular clusters, including essential, 
fluctuating, encounter processes that are thought to operate in such 
clusters, as also those continuous processes which operate in isolated 
binaries and so apply here as well. We can also follow the evolution 
of $N_{XB}$, as also that of the orbital-period distribution of compact 
binaries in globular clusters. For the latter study, however, proper 
modeling of stellar-evolutionary effects still remains to be done for 
parts of the parameter space, as explained in Paper I, and as discussed
in Sec.~\ref{discuss}.   

In Sec.~\ref{revcont}, we briefly review the continuous-limit results
of Paper I, in order to put the results of this paper in their proper 
context. We give only the essentials here, citing figures in Paper I 
for detailed results. In Sec.~\ref{sbe}, we introduce stochasticity
explicitly through our prescription, explaining the details of 
Wiener processes and the It\^{o} calculus in the Appendices. We 
describe our generalization of the Lax-Wendorff scheme, introduced
in Paper I, to handle the solution of the SPDE which the Boltzmann
equation has become now. In Sec.~\ref{res}, we describe the results
of our full calculations including stochasticity, and compare 
these with the continuous-limit results of Paper I. In 
Sec.~\ref{comparison}, we compare our full results with observations.
Finally, In Sec.~\ref{discuss}, we discuss our results, putting them 
in the context of previous studies in the subject, and indicating
some additional physical effects to be included by stages in future 
versions of our scheme, as well as some future problems to be tackled. 

\section{Brief Review of Continuous Limit}
\label{revcont}
  
In order to put the stochastic studies of this paper in their proper
context, we review in this section the essentials of the 
continuous-limit studies of Paper I which form this context. In the   
latter limit, the Boltzmann description works in terms of appropriate
\emph{mean} values of the variables and parameters which are actually
stochastic. Accordingly, the above compact-binary distribution function 
$n(a,t)$ is replaced by its mean value $\nbar(a,t)$, and the Boltzmann
equation has the form:
\begin{equation}
\frac{\partial\nbar(a,t)}{\partial t} = \Rbar(a) - 
\nbar(a,t)\Dbar(a) - \frac{\partial \nbar(a,t)}{\partial a}\fbar(a).
\label{eq:Evol}
\end{equation}
Here, $\Rbar(a)$ is the mean formation rate (per unit binary radius) 
of compact binaries of radius $a$, $\Dbar(a)$ is the mean destruction 
rate per binary and $\fbar(a)$ is the mean shrinkage rate $\dot a$ 
of a compact binary of radius $a$, as described in Paper I. 

\subsection{Mean rates}
\label{revrate}

The mean shrinkage or ``hardening'' rate $\fbar(a)=\dot a$ has been 
given in Fig.~2 of Paper I as a function of $a$, describing the
situation as the compact binary goes from its widest, 
pre-X-ray-binary (PXB) phase to Roche lobe contact, 
and continues through the mass-transferring X-ray-binary (XB) phase.
As shown there, collisional hardening, \ie, that due to encounters 
between the binary and the stellar background of the globular cluster, 
dominates at large $a$, while hardening by gravitational radiation and 
magnetic braking dominates at small $a$. The relative orbit shrinkage 
rate $\dot a/a$ scales roughly as $a$ at large orbital radii, passes 
through a minimum at a critical separation where the gravitational 
radiation shrinkage rate, scaling as $\dot a/a\sim a^{-4}$, takes over.
Magnetic braking also contributes at small radii, but Roche lobe 
contact also occurs at roughly the same point, 
whereupon the angular-momentum transfer 
associated with mass transfer in the XB phase dominates the 
orbit-change rate, and $\dot a$ has a very weak dependence on $a$ 
during this phase. Detailed quantitative expressions for the above
rates are given in Paper I, to which we refer the reader, recording
here only that the orbit-shrinkage rate is given in terms of the 
angular-momentum loss rate by 
\begin{equation}
\frac{\dot a}{a} = 2\frac{\dot J}{J} -2\frac{\dot m_c}{m_c} 
-2\frac{\dot m_X}{m_X},
\label{eq:Ja}
\end{equation}
where $m_X$ is mass of the degenerate star in the compact binary, and 
$m_c$ is that of its (low-mass) companion, and the total 
angular-momentum loss rate can be written in terms of its components    
as:
\begin{equation}
j(a)\equiv\frac{\dot J}{J} = j_{GW}(a) + j_{MB}(a) + j_{coll}(a).
\label{eq:TOT}
\end{equation}
Here, the subscripts `GW', `MB' and `coll' respectively stand for 
gravitational radiation, magnetic braking, and collisional hardening.

The other essential mean rates are those of compact binary formation
and destruction, $\Rbar(a)$ and $\Dbar(a)$, respectively. Consider
first the formation of compact binaries with degenerate primaries 
(white dwarfs or neutron stars) and low-mass companions, such as we are 
interested in this series of papers, in globular cluster (henceforth 
GC) cores. The two relevant dynamical processes are (i) tidal capture 
(tc) of a degenerate, compact star by an ordinary, low-mass star, and 
(ii) an exchange encounter (ex1) between such a compact star and a 
binary of two ordinary low-mass stars in the GC, wherein the compact star 
replaces one of the binary members. Accordingly, the total mean rate 
of formation of compact binaries per unit binary radius, $\Rbar(a)$, 
consists of the above mean tc rate $r_{tc}(a)$ and mean ex1 rate 
$r_{ex1}(a)$:
\begin{equation} 
\Rbar(a)=r_{tc}(a) + r_{ex1}(a).
\label{eq:r_tot}
\end{equation}
The above mean rates are shown as functions of $a$ in Fig.~3 of Paper 
I, and detailed mathematical expressions for them are also given in 
that reference, which we do not repeat here. The mean tidal-capture 
rate is nearly constant for $a<5\Rs$, and decreases rapidly at larger 
$a$. Extensive discussion of various issues related to tidal capture
and of the current status of our understanding of this process are
also given in Paper I, to which we refer the reader. Further 
discussion of previous studies of tidal capture in this problem are
given in Sec.~\ref{discuss}. The mean exchange (ex1) rate is roughly 
constant over the range of radii of interest here, for the 
widely-adopted radius distribution of primordial binaries, \viz, a 
uniform distribution in $\ln a$, which we adopt throughout our work.  

Consider now the destruction of compact binaries with degenerate 
primaries and low-mass companions, which can occur in the following
two ways. First, an encounter with a star which has a relative speed 
higher than an appropriate critical speed can lead to its dissociation 
(dss). Second, in an exchange encounter (ex2) of this binary with a 
compact star, the latter can replace the low-mass companion in the 
binary, forming a double compact-star binary consisting of two neutron 
stars, two white dwarfs, or a neutron star and a white dwarf (all with
masses $m_X\approx 1.4\Ms$ in our model: see Paper I). This destroys 
the binary as an X-ray source (as accretion is not possible in such a 
system), and so takes it out of reckoning in our study\footnote{Note
that it is essentially impossible for one of the compact stars in such 
a double-compact system to be re-exchanged with an ordinary star in a 
subsequent exchange encounter, since the average mass of a background 
GC star (taken as $m_f=0.6\Ms$ in our work) is much less than the 
above value of $m_X$.}. The total mean destruction rate $\Dbar(a)$ 
\emph{per binary} is thus the sum of the above mean dss and ex2 rates:
\begin{equation}
\Dbar(a)=r_{ex2}(a) + r_{dss}(a)
\label{eq:d_tot}
\end{equation}
The above mean rates are shown as functions of $a$ in Fig.~3 of 
Paper I, and detailed mathematical expressions for them are given 
in that paper, which, once again, we do not repeat here.  
The mean dissociation rate is negligible below a critical radius $a_c$ 
corresponding to the above critical speed, rises extremely sharply 
above $a_c$ at first, and eventually scales as $a^2$ for $a\gg a_c$. 
By contrast, the mean exchange (ex2) rate is roughly $\propto a$ and
dominates the destruction processes completely at all orbital radii 
relevant to our study, dissociation becoming important only for very 
soft binaries ($a>1000\Rs$, say), which are of little interest to 
us here.

\subsection{Results}
\label{revresult}

We now summarize the essentials of our continuous-limit results in 
Paper I. We computed evolution of the compact-binary distribution
in this limit, showing the surface $n(a,t)$ explicitly in three 
dimensions in Fig.~5 of Paper I, corresponding to representative   
GC parameters (rather similar to those of the well-known Galactic 
cluster 47 Tuc). The surface evolved smoothly, with the compact-binary 
population growing predominantly at shorter radii ($a<10\Rs$, say). 
We showed that, starting with a small number of binaries at $t=0$ 
following various distributions, we obtained at times $\sim$ Gyr 
or longer a distribution which was independent of the initial 
conditions, determined entirely by the dynamical processes of 
formation and destruction, and by the various hardening processes 
summarized above. We clarified the nature of the distribution and
its evolution by displaying slices through the above surface at 
various points along time axis and $a$-axis, shown in Figs.~6 and 7
of Paper I. The former figure showed that the profile $n(a)$ increased 
with time, roughly preserving its profile for $t>1.5$ Gyr, this profile 
consisting of a roughly uniform distribution at small orbital radii, 
$a\le 6\Rs$, say, and a sharp fall-off at larger radii. The latter
figure showed that $n(a)$ at a given $a$ increased with time and 
approached saturation on a timescale $6-12$ Gyr, the timescale being 
longer at smaller values of $a$. 

For comparison with crucial X-ray observations of Galactic GCs,
we computed the total number of XBs $N_{XB}$ in a GC at any time, 
obtained by integrating $n(a,t)$ over the range of $a$ relevant for 
XBs, \viz, $a_{pm}\le a\le a_L$, where $a_{pm}$ is the value of $a$ 
corresponding to the period minimum $P\approx 80$ minutes, and $a_L$ 
is the value of $a$ at the first Roche lobe contact and onset of 
mass transfer, as explained above: 
\begin{equation}
N_{XB}(t)=\int_{a_{pm}}^{a_L}n(a,t)da.
\label{eq:nxb}
\end{equation} 
Taking an evolutionary time $\sim 8$ Gyr as representative, we 
determined $N_{XB}$ at this point in time, and studied its 
dependence on the Verbunt parameters $\Gamma$ and $\gamma$ that 
describe the essential dynamical properties of globular clusters in 
this context, as explained in Paper I, showing our results as the
computed $N_{XB}(\Gamma,\gamma)$ surface in three dimensions in
Fig.~8 of that paper. 
 
The Verbunt parameters $\Gamma$ and $\gamma$ have been introduced 
in Paper I. Following pioneering suggestions by Verbunt and co-authors
\citep{vh87,v.et.al89}, the crucial importance of these parameters in GC 
dynamics has been lucidly summarized recently by Verbunt 
\citep{v2002,v2006}, and the importance of the parameter $\Gamma$ 
for scaling between different GCs has been emphasized in a pioneering 
study of the production of recycled pulsars in GCs by \citet{dr92}.
Briefly, the first parameter $\Gamma$ is the two-body 
stellar encounter rate, which scales with $\rho^2r_c^3/v_c$, and occurs 
naturally in the rates of all two-body processes, where the standard GC 
core  variables are the average stellar density $\rho$, the velocity 
dispersion $v_c$, and the core radius $r_c$. In fact, we defined $\Gamma$ 
as 
\begin{equation} 
\Gamma\equiv{\rho^2r_c^3\over v_c}\propto\rho^{3/2}r_c^2,
\label{eq:Gamma}
\end{equation}
in Paper I. Note that the last scaling in the above equation holds
only for virialized cores, where the scaling $v_c\propto\rho^{1/2}
r_c$ can be applied. The second parameter is a measure of the rate of 
encounter between binaries and single stars in the GC, the rate normally 
used being the encounter rate $\gamma$ of a \emph{single} binary with 
the stellar background, with the understanding that the total rate of 
binary-single star encounter in the cluster will be $\propto n\gamma$. 
We defined $\gamma$ in Paper I as
\begin{equation} 
\gamma\equiv{\rho\over v_c}\propto\rho^{1/2}r_c^{-1}, 
\label{eq:gamma}
\end{equation}
where the last scaling holds, again, only for virialized cores.  

We compared the results of our above computations with the systematics 
of recent observations of X-ray binaries in Galactic globular clusters 
\citep{pool2003}, as displayed in Figs.~9 and 10 of Paper I.  
We showed that the computed total number $N_{XB}$ of 
XBs expected in a globular cluster scaled in a characteristic way with 
the Verbunt parameters. The qualitative nature of this scaling was 
rather similar to that found in our earlier ``toy'' model 
\citep{bg2006}, although details were different. $N_{XB}$ scaled with 
$\Gamma$ (which is a measure of the dynamical formation rate of compact 
binaries, as above) and, at a given $\Gamma$,  $N_{XB}$ decreased with 
increasing $\gamma$ (which is a measure of the rate of destruction of these 
binaries by dynamical processes) at large values of $\gamma$ , as
shown in Fig.~9 of Paper I. This rough scaling could be expressed as 
$N_{XB}\propto\Gamma g(\gamma)$, where the ``universal'' function 
$g(\gamma)$ of $\gamma$ (except for a spurious feature at low values of 
$\gamma$ which we explained in Paper I) decreased monotonically with 
increasing $\gamma$, reflecting the increasing strength of dynamical 
binary-destruction processes with increasing $\gamma$. 

We further demonstrated that these computed trends with the Verbunt 
parameters compared very well with the observed trends in above X-ray 
data by showing in Fig.~10 of Paper I the contours of constant $N_{XB}$ 
in the $\Gamma-\gamma$ plane, and overplotting on the positions of the
Galactic GCs with observed X-ray binaries in them. We showed that the 
trend in the observed $N_{XB}$ values generally followed the contours, 
with one exception. This provided us with a first indication of the basic 
ways in which dynamical binary formation and destruction processes work 
in GCs, and encouraged us to build more ``realistic'' models by introducing 
stochastic effects explicitly, to which this paper is devoted.   

\section{Introducing Stochasticity}
\label{sbe}

In order to study the behavior of the inherently stochastic terms 
in the full Boltzmann equation
\begin{equation}
\frac{\partial n(a,t)}{\partial t} = R(a,t) - n(a,t)D(a,t) - 
\frac{\partial n(a,t)}{\partial a}f(a,t),
\label{eq:Evol_s}
\end{equation}
we must explicitly include stochastic, fluctuating parts in these 
terms, in addition to their mean values studied in Paper I, as above.   
We do so by expressing the above rates $R(a,t)$, $D(a,t)$, and 
$f(a,t)$ as their earlier mean values $\Rbar(a)$, $\Dbar(a)$ and 
$\fbar(a)$, augmented by fluctuating components as below: 
\begin{equation}
\left .
\begin{array}{l}
R(a,t)=\Rbar(a)+\zat_{tc}+\zat_{ex1}\\
D(a,t)=\Dbar(a)+\zat_{ex2}+\zat_{dss}\\
f(a,t)=\fbar(a)+\zat_{coll}\\
\end{array} 
\right\rbrace 
\label{eq:fluc}
\end{equation}
Here, $\zat_{X}$ is the random fluctuation rate of events of type X
from their mean rates, and X = tc, ex1, ex2, dss, coll by turn, these 
notations having been introduced above. In general, $\zat_{X}$ is a 
function of both $a$ and $t$, of course.   

The crucial question is that of modeling $\zat_{X}$ appropriately. In
this introductory work, we use the standard normally-distributed model
\begin{equation}
\zat_X=S_X(a)\eta^t,
\label{eq:zeta}
\end{equation}
where $S_X^2(a)$ is the variance of $\zat_X$ at a given $a$ and 
$\eta^t$s at each $t$ are independent random numbers distributed 
in a standard normal distribution. This separable form is appropriate 
since the dynamical processes of binary formation and destruction at a 
given value of $a$ are inherently independent of those at other values 
of $a$. The ``flow'' or ``current'' of binaries from larger to smaller 
values of $a$ due to the hardening described above and in Paper I 
does not affect this independence, but merely changes the number 
of binaries in an infinitesimal interval of $a$ around a given value 
of $a$ at a given instant $t$, which is automatically taken into 
account by the Boltzmann equation (also see below). 
Indeed, the hardening process itself has this independence, \viz, that 
its rate at a given value of $a$ is independent of that at other values 
of $a$, and so is separable in the same way. By contrast, the number 
distribution $n(a,t)$ of the binaries  \emph{cannot} be written in 
this form, since, at a particular $a$, it is determined both by the 
binary formation and destruction rates at that $a$, \emph{and} by 
the rates of binary arrival from (and also departure to) other values 
of $a$ due to hardening, as described above. All of this is, of course,
automatically included in the Boltzmann equation by definition.

The essence of the physics of these fluctuations is contained in the 
adopted model for $\eta^t$. By adopting a normally-distributed 
variation, we are, in effect, considering a \emph{Wiener process}
(see Appendix A and references therein), which is the standard
mathematical description of Brownian motion. In other words, we are 
studying a situation wherein the variations in the above dynamical 
rates about their respective mean values constitute a Brownian 
motion. We return to Wiener processes later in more detail. 

\subsection{Variances of stochastic-process rates}
\label{var}

How do we estimate the variance of a stochastic process of type X
whose mean value is $\Rbar_X(a)$? To answer this question, consider 
first how it is addressed in Monte Carlo simulations, which have 
been performed in this subject by several authors ( see, \eg, 
 \citet{sp93}, \citet{z197}, or \citet{fr2003}).  These works have 
uniformly used the so-called \emph{rejection method} for 
determining whether an event of a given type occurs in a given 
time interval or not. The method works as follows.

For events of type X, if the mean rate of event occurrence is 
$\Rbar_X$, then the timescale for occurrence of such events is
\begin{equation}
{\Delta t}_X=\frac{1}{\Rbar_X}
\label{eq:tscale}
\end{equation}
Hence, during a time step $\Delta t < {\Delta t}_X$, the quantity 
$p_X=\Rbar_X\Delta t<1$ is the \emph{expected mean number 
of events} during this interval. $p_X<1$ can also be interpreted as 
the \emph{probability} of occurrence of an event X within this time
step \citep{z197}, and the actual number of such events within 
$\Delta t$ will then follow a binomial distribution with the following 
mean and variance:
\begin{equation}
\left .
\begin{array}{rl}
{\rm mean}=	&	\Rbar_X(a)\Delta t\\
{\rm variance}=S_X^2(a)\Delta t^2=	&	
\Rbar_X(a)\Delta t(1-\Rbar_X(a)\Delta t).
\end{array}
\right \rbrace
\label{eq:mv}
\end{equation} 
Note that the above variance depends on $a$, since the mean rates 
depend on $a$. When several different types of events are considered 
simultaneously, as in the present problem, we must, of course, so 
choose $\Delta t$ that it is shorter than the shortest event-occurrence
timescale appearing in the problem. We discuss this point below. 

\subsubsection{Time step}
\label{tstep}

The mean rates depend on $a$ as detailed in Paper I (see Fig.~3 of
that paper). $\Rbar_{tc}(a)$ is a decreasing function of $a$, and 
so attains its maximum at $a=a_{min}$. All other rates are either 
constant (ex2), or increasing functions of $a$, so that their 
maximum values can be thought to occur at $a=a_{max}$. Accordingly, 
if we make the following choice for our computational time step 
$\Delta t_d$:
\begin{equation}
\Delta t_d<{\rm min}\left \lbrace
{1\over\Rbar_{tc}(a_{min})},{1\over\Rbar_{ex1}(a_{max})},
{1\over\Rbar_{ex2}(a_{max})},{1\over\Rbar_{dss}(a_{max})},
{1 \over \overline{\dot a_{coll}}(a_{max})} \right \rbrace,
\label{eq:dt_d}
\end{equation}
this will ensure that $\Delta t_d$ is smaller than the shortest of
the above event-occurrence timescales. 

However, as is well-known, this time step must also obey the 
\emph{Courant condition} \citep{pr1992} throughout the range of $a$ 
under consideration (\ie, 0.6\Rs-60\Rs):
\begin{equation}
\Delta t_c=\epsilon \frac{\Delta a}{\fbar_{\rm max}}, 
\qquad \epsilon<1.
\label{eq:dt_c}
\end{equation}
Here, $\Delta a$ is the step-size in $a$, and $\fbar_{\rm max}$ is the 
largest value of $\fbar(a)$ over the range of $a$ under consideration (see 
above and paper I). Satisfaction of this condition is essential for the 
stability \citep{pr1992} of the solution of Eqn.~(\ref{eq:Evol_s}).
 
To ensure that both of the above conditions are satisfied, we choose 
the time step $\Delta t$ for solving Eqn.~(\ref{eq:Evol_s}) to be
\begin{equation}
\Delta t={\rm min}\lbrace \Delta t_d, \Delta t_c \rbrace .
\label{eq:dt}
\end{equation} 

\subsection{Solution of Stochastic Boltzmann Equation}\label{soln}

The Lax-Wendorff scheme \citep{pr1992} used by us for numerical 
solution of the Boltzmann equation in the continuous limit has been 
introduced in Sec.~2.6 of Paper I. The stochastic version of this equation, 
\viz, Eqn.~(\ref{eq:Evol_s}) can be looked upon as the earlier continuous 
equation with additional stochastic terms, which turns it into a 
SPDE (see Sec.~\ref{intro}). We now discuss our method of solving
this SPDE\footnote{In SPDE literature, the continuous terms are
sometimes called \emph{drift} terms and the stochastic terms 
\emph{diffusion} terms, but we shall \emph{not} use this 
terminology here, since stochastic terms in our problem do not 
always represent diffusion, and furthermore since there is a 
possibility with such usage of confusion with the Fokker-Planck 
approach, which does represent diffusion in phase space.}. 
 
It it well-known that ordinary calculus cannot be applied to the 
handling of stochastic terms in SPDEs, since these terms are
non-differentiable in the ordinary sense, and the ordinary 
definition of an integral does not apply to them. Rather, one has
to modify the methods of calculus suitably, and redefine appropriate 
integrals. As summarized in Appendix B, one such modified calculus 
is the $It \hat o{\rm~}Calculus$, which has been used widely for 
solution of SPDEs in recent years \citep{ok2004,kl94}. The 
corresponding integrals involving the stochastic terms are then 
called $It\hat o{\rm~}integrals$, which have properties appropriately 
different from those of ordinary integrals, as indicated in Appendix B. 

\subsubsection{Numerical Method}\label{num}

In solving an SPDE like Eqn.~(\ref{eq:Evol_s}), one integrates
the continuous terms in the usual way, but the stochastic terms 
must be integrated using It$\hat o$ calculus \citep{g95}. This means 
that, in advancing the solution at $t$ by a time step $dt$ --- which is 
essentially a Taylor expansion of the solution $n(a,t)$ about $t$ --- 
the expansions of the stochastic terms in Eqn.~(\ref{eq:Evol_s}) are 
to be performed using the stochastic Taylor expansion 
(Eqn.~(\ref{eq:taylorexp})), as discussed in Appendix B. 
 
A variety of numerical algorithms have been explored by various 
authors for numerical solution of SPDEs. The particular algorithm we 
use is a hybridization of the two-step Lax-Wendorff scheme for the 
continuous terms, as utilized in paper I, and the second-order stochastic 
Taylor expansion according to the \emph{Milshtein scheme} for the 
stochastic terms \citep{m74,g95}, \ie, Eqn.~(\ref{eq:taylor2}), as 
explained in Appendix B. In this scheme, there is only one stochastic 
path to be solved for in our case \viz, that of $n(a,t)$ (corresponding 
to $X_k$) and the continuous terms (\ie, the $\sigma_p$s), the 
variances in tc, ex1, ex2, dss and coll rates being as given above. 
Note that, in each of the two steps in the Lax-Wendorff scheme, the 
expansion (\ref{eq:taylor2}) needs to be applied, whereupon we arrive 
at the following discretization scheme\footnote{It can be shown that 
the commutation condition (\ref{eq:comm}) is satisfied in this case.} 
for Eqn.~(\ref{eq:Evol_s}):
\begin{equation}
\begin{array}{l}
{\rm~Half~step:}\\
n_{j+1/2}^{N+1/2}={1\over2}\left(n_{j+1}^N+n_j^N\right)
+\left[\Rbar(a_{j+1/2}) - \Dbar(a_{j+1/2})\left(\frac{n_{j+1}
^N+n_j^N}{2}\right)\right]{\Delta t\over 2} \\
+ \left( \Whan_{tc} + \Whan_{ex1} \right) -
\left( \Whan_{ex2} + \Whan_{dss} \right)
\left(\frac{n_{j+1}^N+n_j^N}{2}\right)\\
+ \left[\left((\Whan_{ex2})^2 -S_{ex2}^2(a_{j+1/2})\right) + 
\left((\Whan_{dss})^2 -S_{dss}^2(a_{j+1/2})\right)\right]
\left(\frac{n_{j+1}^N+n_j^N}{4}\right)\\
+ \left(\Whan_{ex2}\Whan_{dss}\right)
\left(\frac{n_{j+1}^N+n_j^N}{2}\right)\\
-\frac{\fbar(a_{j+1/2})\Delta t}{2\Delta a}(n_{j+1}^N-n_j^N)
-\frac{\Whan_{coll}}{2\Delta a}(n_{j+1}^N-n_j^N),\\               
{\rm~Full~step:}\\              
n_j^{N+1} = n_j^N + \left(\Rbar(a_j)-\Dbar(a_j)n_j^N\right)
\Delta t\\ + \left( \Wan_{tc} + \Wan_{ex1} \right) -
\left( \Wan_{ex2} + \Wan_{dss} \right)n_j^N\\
+ \left[\left((\Wan_{ex2})^2 -S_{ex2}^2(a_j)\right) + 
\left((\Wan_{dss})^2 -S_{dss}^2(a_j)\right)\right]
\frac{n_j^N}{2}\\
+\left(\Wan_{ex2}\Wan_{dss}\right)n_j^N\\
-\frac{\fbar(a_j)\Delta t}{\Delta a}\left(n_{j+1/2}^{N+1/2}
-n_{j-1/2}^{N+1/2}\right)-\frac{\Wan_{coll}}{\Delta a}
\left(n_{j+1/2}^{N+1/2}-n_{j-1/2}^{N+1/2}\right).
\end{array}
\label{eq:lwm}
\end{equation}
Here, $\Wan_{X}\equiv S_X(a_j)\eta^N\Delta t$, where $\eta^N$ is
the value of a standard normal variate at the $N$th time step. 

For any particular run, we compute the $\Wan_X$s ($\Whan_X$s) for 
a particular $a_j$ ($a_{j+1/2}$) over the $a$ and $t$ intervals of 
integration, and repeat it for all $a_j$s. 
The standard normal variate $\eta^N$s are generated using the 
well-known \emph{polar method} \citep{pr1992}. All values of 
$\Wan_X$ and $\Whan_X$ are stored in a two dimensional array 
(\ie, a \emph{Wiener sheet}), which serves as the input for solving
Eqn.~(\ref{eq:lwm}). Because of the fluctuations in the collisional 
hardening rate (as contained in $\zat_{coll}$), it is not impossible 
that the value of the total hardening rate $f$ might occasionally
exceed $\fbar_{max}$, which would violate the Courant condition, 
possibly making the solution procedure unstable. To avoid this, we 
have so restricted the variations in $\Wan_{coll}$s and 
$\Whan_{coll}$s that the \emph{amplification factor} $\epsilon 
\equiv f\Delta t/\Delta a$ always lies between zero and unity 
\citep{pr1992}. 

\section{Results}
\label{res}

We now present the results obtained from our above computations
of the cases which we studied in Paper I in the continuous limit.
As before, we study (a) the evolution of the distribution function 
$n(a,t)$, and, (b) the dependence of the computed number of XBs 
$N_{XB}$ on the Verbunt parameters. We choose exactly the same 
values of all GC parameters as we did in Paper I, for ease of 
comparison. 

\subsection{Evolution of compact-binary distribution}\label{cbdist}

\begin{figure}
\plotone{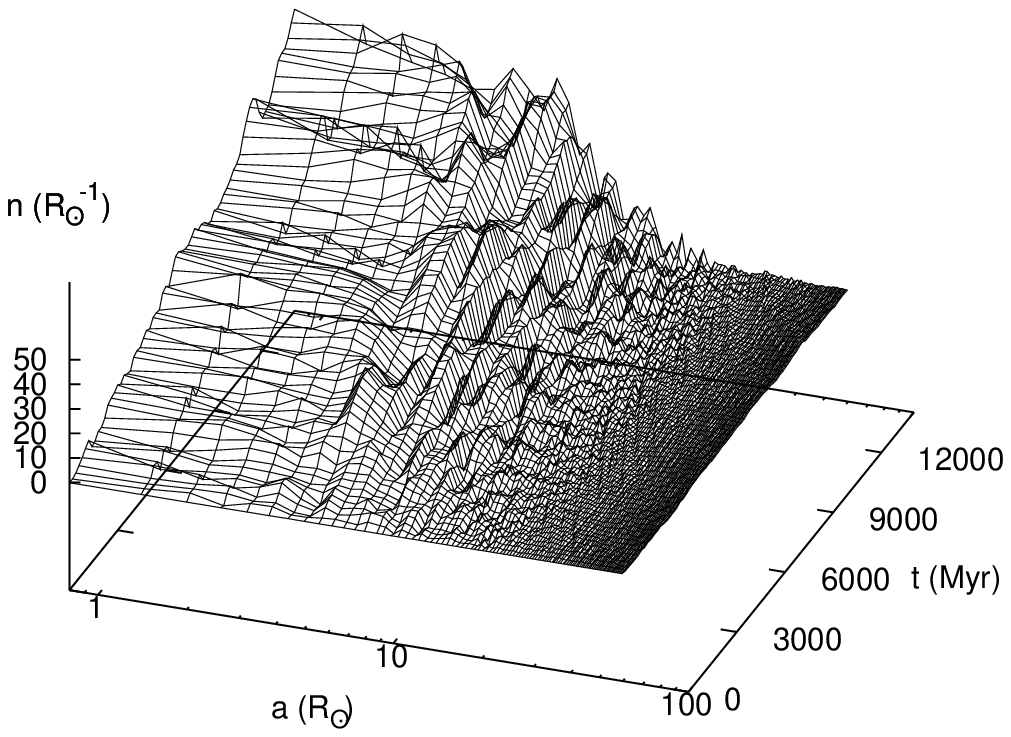}
\caption{A typical example, \ie, one ``realization''of the evolution of 
the binary distribution function $n(a,t)$. Globular cluster parameters 
are chosen to be roughly those of 47 Tuc, as explained in text (also 
see Fig.~5 of Paper I).}
\label{fig:ev3d}
\end{figure}

In Fig.~\ref{fig:ev3d}, we show a typical evolution of the compact 
binary population distribution $n(a,t)$. The GC parameters were 
chosen, as in Paper I, to be $\rho=6.4\times10^4 \Ms {\rm pc}^{-3}$,
$r_c=0.5{\rm~pc}$ and $v=11.6{\rm ~km}{\rm ~sec}^{-1}$, similar 
to those of the well-known Galactic cluster 47 Tuc. As the figure 
shows, the surface representing the evolution fluctuates randomly 
throughout, but it does show a clear overall evolution which is of the 
same nature as that in the continuous limit (cf. Fig.~5 of Paper I). 
In particular, the population grows with time predominantly at 
shorter radii ($a<10\Rs$). As before, we start with a small number 
of primordial compact binaries with various initial distributions, 
and find that, by $t \sim 1-1.5$ Gyr, the distribution ``heals'' to a 
form which is independent of the initial choice of distribution. 
The fluctuations differ in detail from run to run, of course, as
we choose different seeds for random number generation, but the 
overall nature of the evolution remains the same for all runs. Indeed, 
the results for different runs seem to represent different variations 
about a \emph{mean} surface, which is very close to that in the 
continuous limit, as given in Paper I. We explicitly demonstrate this
below by displaying temporal and radial slices through the above 
surface $n(a,t)$ (see Paper I) for different runs, and also displaying 
their averages over a number of runs, which we show to be close to 
the continuous limit.     
    
\begin{figure}
\plotone{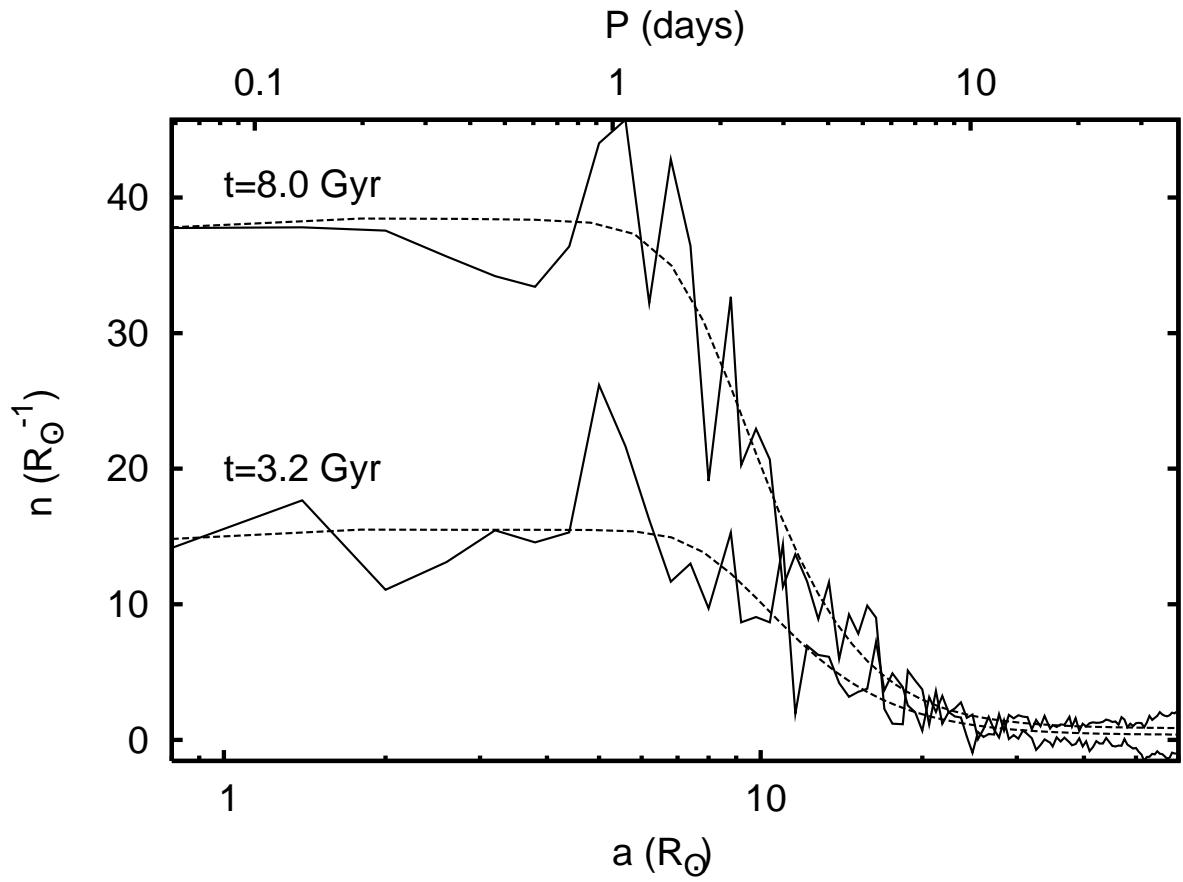}
\caption{Typical time slices, \ie, $n(a)$ at specified times, for the 
evolution shown in Fig.~\ref{fig:ev3d} (solid lines). Overplotted are 
the same time slices in the continuous limit (dashed lines) from
Paper I.} 
\label{fig:snap}
\end{figure}

\begin{figure}
\plotone{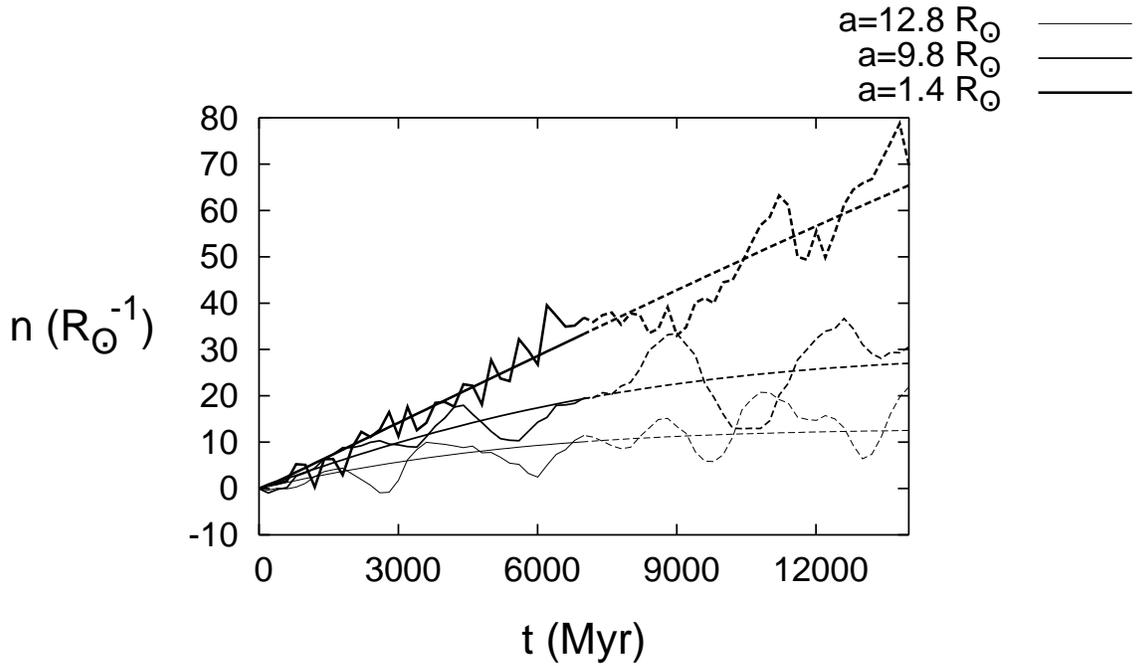}
\caption{Typical radial slices, \ie, $n(t)$ at fixed values of binary 
radius for the evolution shown in Fig.~\ref{fig:ev3d}. Overplotted are 
the same radial slices in the continuous limit from
Paper I. As in Paper I, we show the evolution beyond 8 Gyr by dashed 
lines to indicate that such long evolution times may not be applicable 
to Galactic GC, but are included here to demonstrate the timescales.}
\label{fig:tvol}
\end{figure}

To do this, we first show in Fig.~\ref{fig:snap} typical time 
slices, \ie, $n(a)$ at fixed $t$, (solid lines) through the 
surface in Fig.~\ref{fig:ev3d}, for a \emph{single} run, 
overplotting the continuous limit from Paper I for comparison.
The distribution with fluctuations does indeed follow the 
continuous-limit distribution generally, the same gross features 
being visible through fluctuations, in particular that $n(a)$ is
roughly constant $a\leq 7\Rs$, and falls off sharply at larger  
radii. The overall nearly-self-similar evolution at large
times, described in Paper I, can also be vaguely discerned 
through the fluctuations. We have discussed possible causes of
such self-similar evolution in Paper I. Next, in Fig.~\ref{fig:tvol},
we show radial slices corresponding to the evolution in 
Fig.~\ref{fig:ev3d}, representing the behavior of $n(t)$ at a 
fixed radius $a$, overplotted with the continuous limit from
Paper I. Again, the curves from a single run follow, in a 
statistical sense, the corresponding continuous limits. In 
particular, it can be seen that the radial slices corresponding 
to larger values of $a$ tend to saturate by about 6 Gyr, while 
those for smaller values of $a$ do not show such saturation.  

\begin{figure}
\plotone{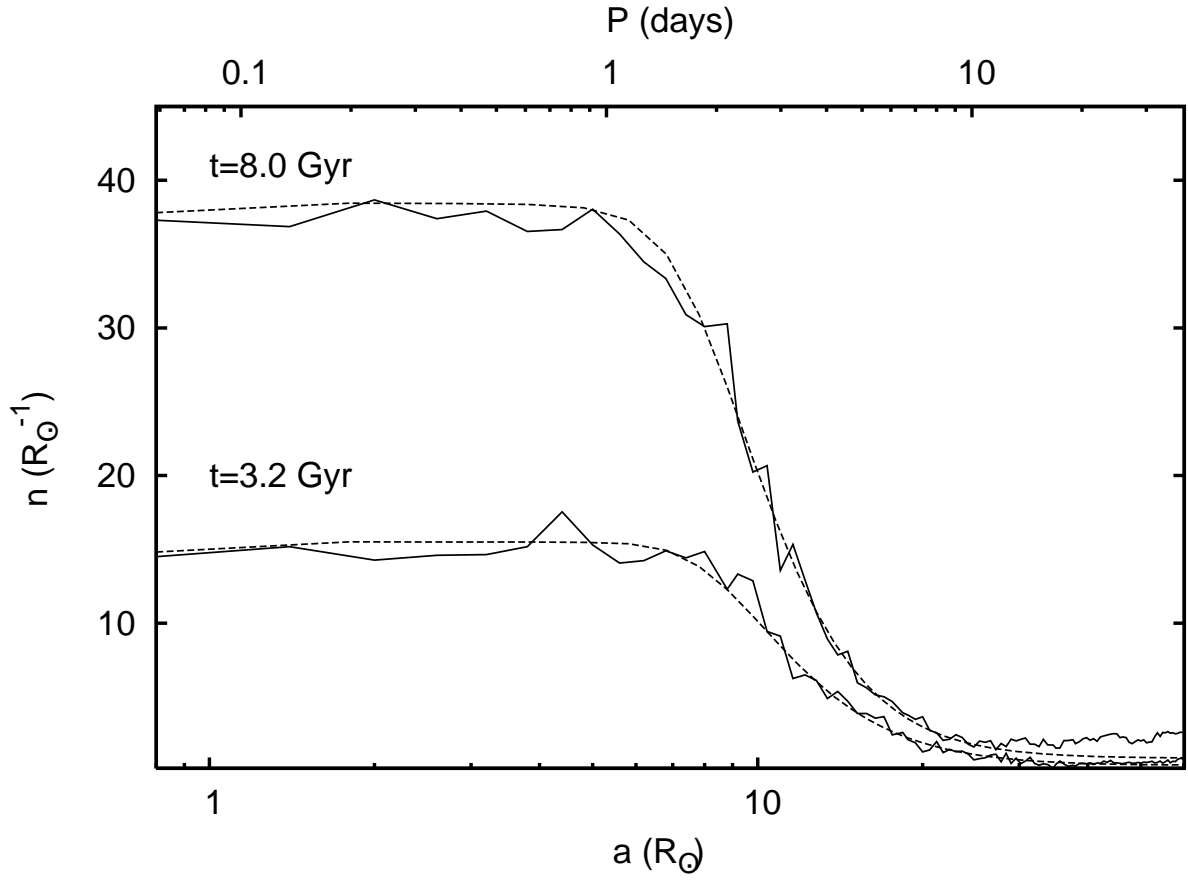}
\caption{Typical time slices through the average evolutionary surface of 12 different 
``realizations'' of the evolution represented in Fig.~\ref{fig:ev3d}, all with the
same GC parameters (solid line). Overplotted are the corresponding time slices in the
continuous limit from paper I (dashed line).} 
\label{fig:snap_avr}
\end{figure}

\begin{figure}
\plotone{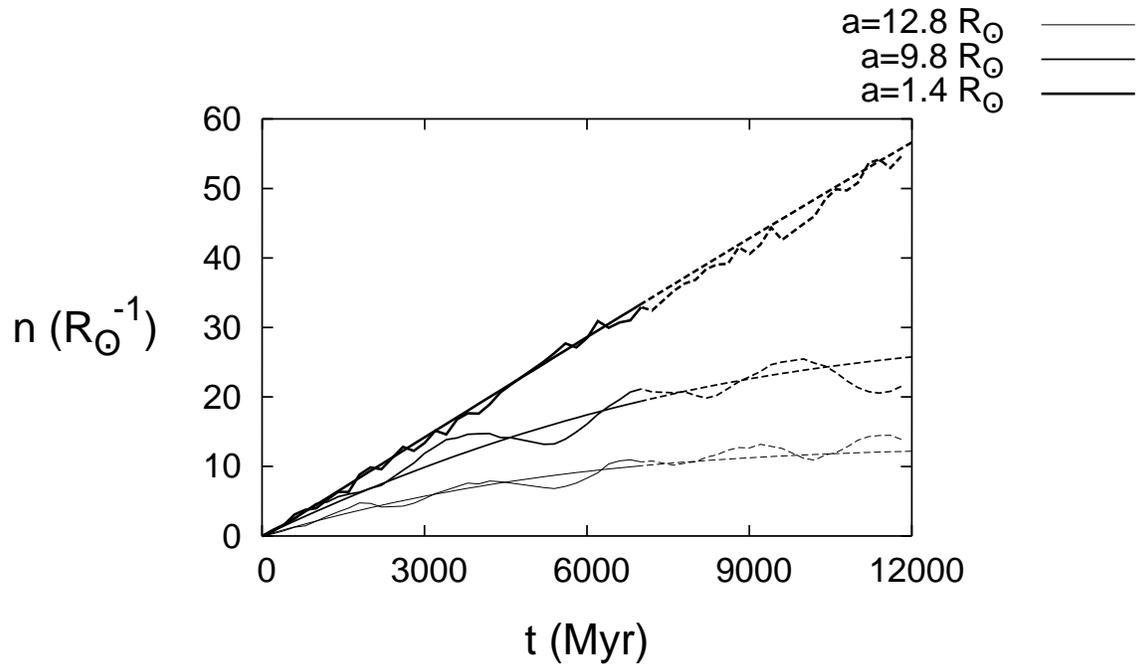}
\caption{Typical radial slices of the same average evolutionary surface as in Fig.~\ref{fig:snap_avr}. 
Overplotted are the corresponding radial slices in the continuous limit from paper I.}
\label{fig:tvol_avr}
\end{figure}

Finally, in Figs.~\ref{fig:snap_avr} and \ref{fig:tvol_avr}, we 
show the above temporal and radial slices of the average of 12 
different runs, overplotted with the the corresponding continuous 
limits. These figures clearly demonstrate how the fluctuations 
average out over many runs, so that the mean result approaches 
the continuous limit.    

\subsection{Number of X-ray binaries}\label{xbnumber}

The total number of GC X-ray binaries $N_{XB}$ at a particular time
was computed from Eq.~(\ref{eq:nxb}), as in Paper I. We determined 
$N_{XB}$ for a representative evolution time of $\sim 8$ Gyr, and 
studied its dependence on the Verbunt parameters $\Gamma$ and 
$\gamma$, so as to relate our computational results with the 
systematics of recent observations of X-ray binaries in globular 
clusters \citep{pool2003}. For this, we computed, as in Paper I, 
values of $N_{XB}$ over a rectangular grid in $\Gamma-\gamma$ space, 
spanning the range $\gamma=1-10^6$ and $\Gamma=10^3-10^8$, which 
encompasses the entire range of Verbunt parameters over which 
Galactic GCs have been observed (see Paper I). Although the GCs 
actually observed so far lie along a diagonal patch over this grid,
as explained in Paper I, computational results over the whole grid 
are useful for clarifying the theoretically expected trends.

As explained in Paper I, at a specific grid point ($\Gamma$, 
$\gamma$), the values of $\rho$, $r_c$ and $v_c$ are evaluated 
using the definitions of Verbunt parameters and the virialization 
condition (see Sec.~3.2 of paper I for a detailed discussion). 
Also as before, we take representative values of primordial stellar 
binary fraction ($k_b$) and compact-star fraction ($k_X$) to be 10 
percent and 5 percent respectively.  

\begin{figure}
\plotone{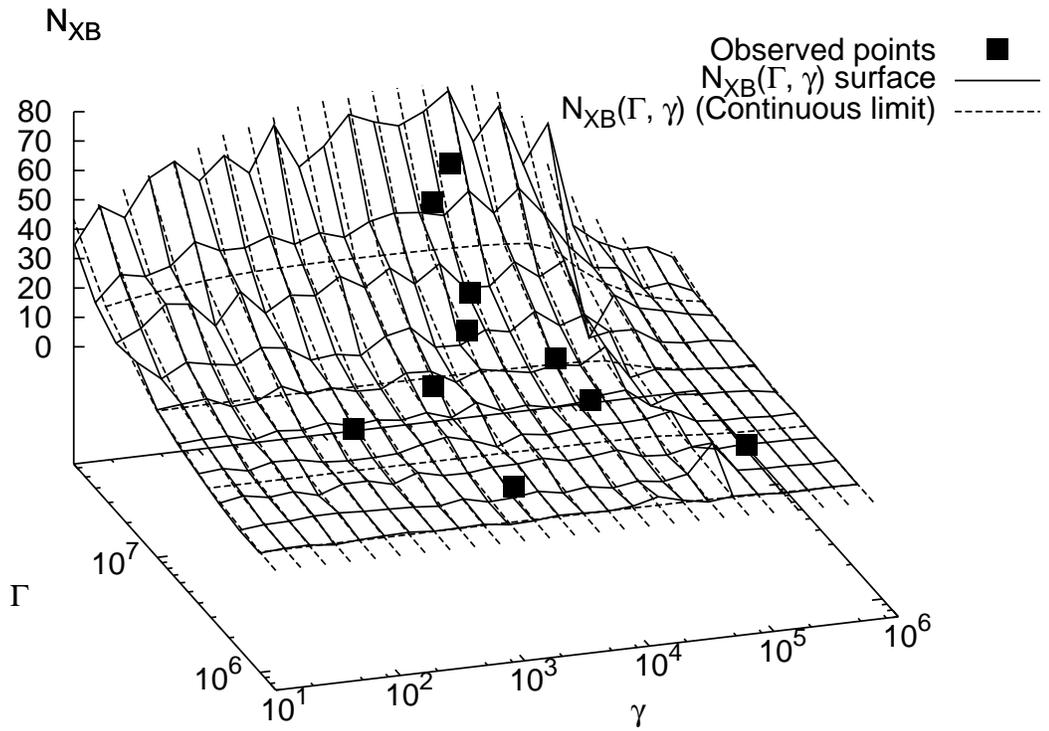}
\caption{$N_{XB}(\gamma, \Gamma)$ surface (solid line). The observed
GCs with significant number of XBs are shown overplotted. Also shown 
overplotted is the continuous-limit result from Paper I (dashed line).}
\label{fig:nxb}
\end{figure}

Fig.~\ref{fig:nxb} shows the resulting $N_{XB}(\Gamma,\gamma)$ 
surface. As indicated in Paper I, the overall fall-off in this 
surface for $\gamma>3\times10^3$ is a signature of the increasing 
rates of compact-binary destruction rates with increasing $\gamma$,
and the above specific value of $\gamma$ represents an estimate
of the threshold above which destruction rates are very important.
Further, the trend in $N_{XB}$ with $\Gamma$ is simple --- 
$N_{XB}$ increases with $\Gamma$ monotonically, since the dynamical
formation rate of compact binaries scales with $\Gamma$. What we 
notice in fig.~\ref{fig:nxb} is that this surface also shows random 
fluctuations due to the stochastic processes, but it generally 
follows the $N_{XB}$ surface corresponding to the continuous limit, 
shown overplotted in the same figure. This is similar to what
was discussed above for the compact-binary distribution, and the 
point about the mean surface corresponding to the average of many 
realizations of the stochastic processes being very close to the 
continuous limit also holds here. We also note that the \emph{total} 
fluctuations in $N_{XB}$ increase with increasing value of $\Gamma$. 
However, as will become evident from results discussed below, the 
\emph{relative} fluctuations actually decrease with increasing 
$\Gamma$.

\begin{figure}
\plotone{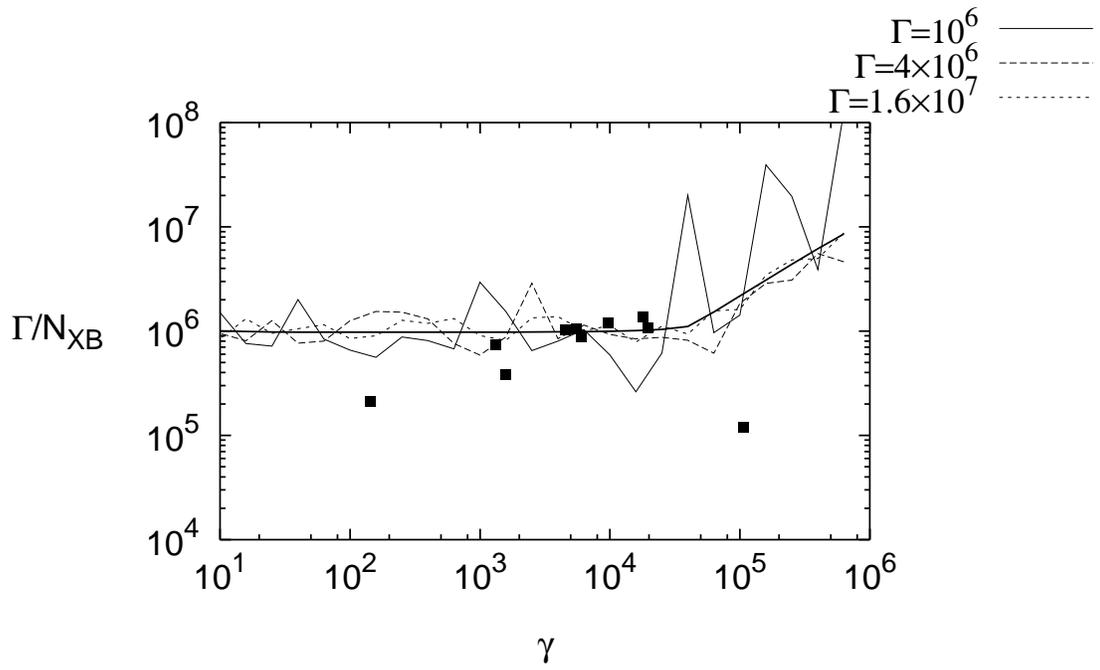}
\caption{Computed $\Gamma/N_{XB}$ as a function of $\gamma$, for 
values of $\Gamma$ as indicated. The continuous-limit result for 
$\Gamma=10^7$ is shown overplotted (thick line). Also shown
overplotted are the positions of Galctic GCs with significant
numbers of X-ray sources, as in Paper I.}
\label{fig:toylike}
\end{figure}

To further clarify the trends and to make comparisons with the
results of the ``toy'' model of B06 and with those of Paper I, we 
plot the quantity $\Gamma/N_{XB}$ for a \emph{fixed} value of 
$\Gamma$ against $\gamma$ in Fig.~\ref{fig:toylike}, displaying
the curves for several values of $\Gamma$ as indicated. As can be 
seen, the fluctuating $\Gamma/N_{XB}$ vs. $\gamma$ curves for
various values of $\Gamma$ follow the same mean trend, although
the details of the fluctuation are different in different cases.
This mean trend is in fact very close to the mean ``universal''
curve corresponding in the continuous limit evolution of Paper I,
and is overplotted in the figure. Thus, as in the continuous limit 
case, the basic scaling of the toy model, \viz, $N_{XB}\propto\Gamma 
g(\gamma)$, where $g(\gamma)$ is a ``universal'' decreasing function 
(representing the increasing binary destruction rate with increasing 
$\gamma$, as explained above), does essentially carry over to this 
detailed model with stochasticity included, suggesting a robust 
feature of the scaling between different clusters which is expected
to be further confirmed by future observations. 

Another feature of Fig.~\ref{fig:toylike} is that the relative
fluctuations in the curves increase with decreasing value of 
$\Gamma$. This is consistent with the intuitive notion that, in all 
phenomena of this nature, the relative fluctuations in $N_{XB}$ are 
expected to increase at smaller values of $N_{XB}$, which occur at 
smaller values of $\Gamma$. More formally, this can be seen as 
follows. From Eqn.~(\ref{eq:mv}), it is clear that, over an interval 
$\Delta t$, the relative variance in the number of events of type 
X is: 
$$r_X(a)=(1-\Rbar_X(a)\Delta t).$$ 
For the range of $\Gamma$ and $\gamma$ considered in this work, we
found that $\Delta t$ was actually close to $\Delta t_c$ in most 
cases, so that $\Delta t\sim \gamma^{-1}$ roughly. Since the 
formation rates scale as $R_X\sim\Gamma$, we have:  
$$r_X(a)=\left(1-{\mathcal O}{\Gamma \over \gamma}\right).$$ 
Therefore, for a fixed $\gamma$, $r_X(a)$ increases as $\Gamma$ 
(and hence $N_{XB}$) decreases.

\subsection{Comparison with observations}\label{comparison}

In Secs.~\ref{cbdist} and \ref{xbnumber} we saw that the basic
trends of the results, as obtained from the stochastic Boltzmann
equation (\ref{eq:Evol_s}), are the same as those obtained
from the Boltzmann equation in the continuous limit in Paper I.
Therefore, as in paper I, the results from the stochastic Boltzmann
equation are consistent with the observations of XB populations in 
Galactic GCs. Indeed, since fluctuations \emph{are} present in
the dynamical processes under study here, we should ideally  
compare theoretical trends including fluctuations with observational 
results, as we do in this paper, where Fig.~\ref{fig:nxb} shows the 
positions of the observed GCs with significant numbers of X-ray 
sources in the $\gamma-\Gamma-N_{XB}$ co-ordinates. The observational
points do lie near the computed $N_{XB}(\gamma,\Gamma)$ surface. In 
Fig.~\ref{fig:toylike}, we compare the $\Gamma/N_{XB}-\gamma$ curves 
with the positions of the observed points, showing that most points 
do indeed lie near the curves. 

\begin{figure}
\plotone{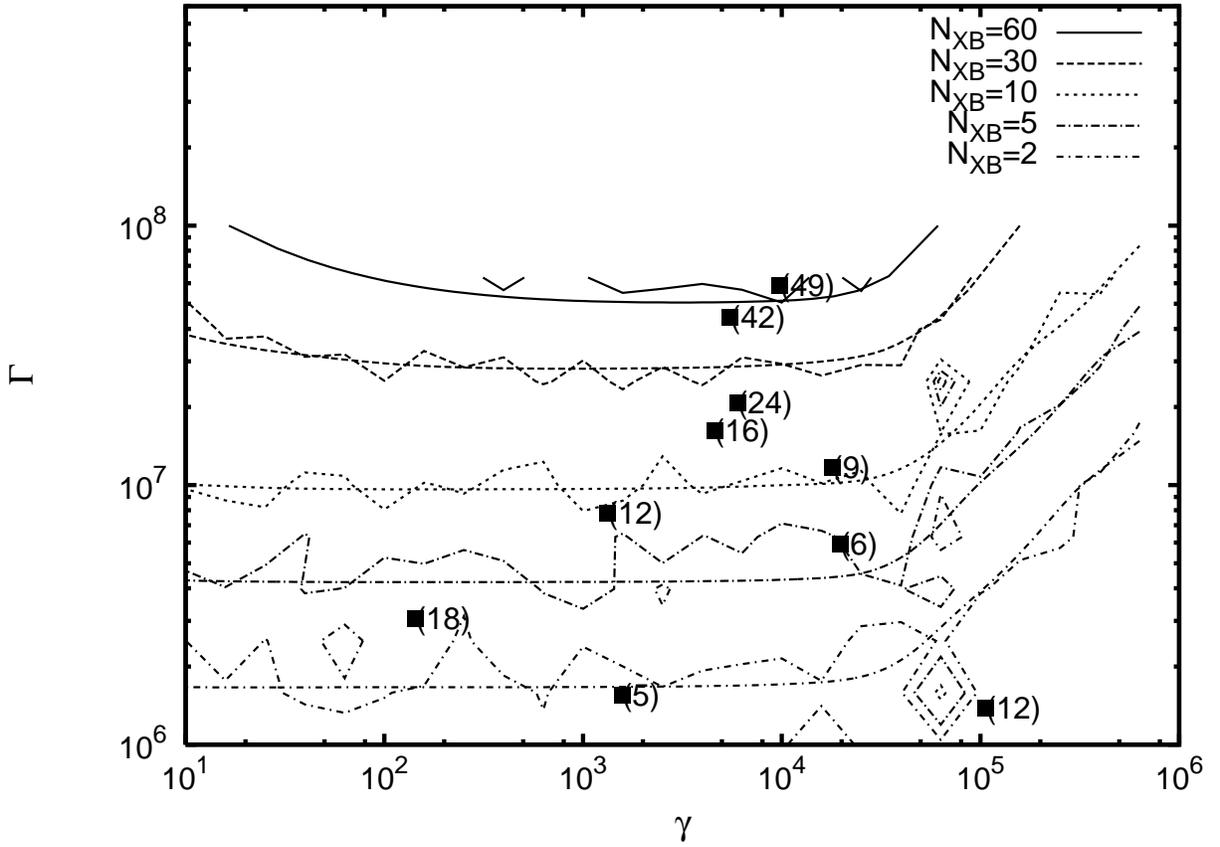}
\caption{Contours of constant $N_{XB}$ in the plane of Verbunt 
parameters. Corresponding contours in the continuous-limit case
are shown overplotted, using the same line-styles for easy 
comparison. Positions of GCs with significant numbers of X-ray 
sources are also overplotted, with the corresponding $N_{XB}$ in 
parentheses, as in Paper I.}   
\label{fig:cntr}
\end{figure}

In Fig.~\ref{fig:cntr} we plot the computed contours of constant 
$N_{XB}$ in the plane of Verbunt parameters, similar to what we did
in Fig.~10 of Paper I, but now with the fluctuations included. The
fluctuations are clearly seen to be larger for smaller values of 
$N_{XB}$, as expected, and as mentioned above. Again, the observed
numbers generally agree well with the present contours which include
fluctuations, and these contours do generally follow the 
continuous-limit contours of Paper I, which are shown overplotted. 

\section{Discussions}\label{discuss}

We have described in this paper a scheme for introducing 
stochasticity into the Boltzmann study of compact-binary evolution 
in globular clusters that we began in Paper I. Our scheme involves 
the use of stochastic calculus (for the first time in this subject, 
to the best of our knowledge), whereas previous studies in the 
subject have normally used Monte-Carlo methods of various 
descriptions --- depending on the particular aspect of the problem 
being studied --- for handling stochasticity (see, \eg, 
\citet{H.et.al92,dr94,fr2003,fr2007}). With the aid of this scheme, 
we have demonstrated that the joint action of inherently stochastic 
and continuous processes produces evolutionary trends which 
necessarily contain fluctuations that vary between individual 
``realizations'' of the stochastic processes, as expected. 
However, these trends do generally follow those found 
in the continuous-limit approximation of Paper I, and when trends
are averaged over more and more realizations, the mean trend comes
closer and closer to the continuous-limit one. In this sense, the 
continuous limit is very useful as an indicator of the expected 
mean trend. On the other hand, the magnitude of the fluctuations 
seen in any given realization, particularly in certain parts of 
parameter space, suggest that one should compare the results of a 
typical realization to observations, in order to get a feel for 
expected fluctuations in the data from stochastic dynamical 
processes alone, \ie, apart from those coming from uncertainties 
in the observational methods of obtaining the data.

Boltzmann approach in its original form appealed to us because 
of its ability by definition to handle weak, frequent, distant 
encounters and strong, rare, close encounters on the same footing. 
Of course, the approach is of practical use only when probabilities
or cross-sections of such encounters are known from detailed 
studies of individual encounters through numerical experiments, as 
is the case for our current use of this approach. It was generally 
believed that, since Fokker-Planck methods were normally used for
handling only the weak, frequent, distant encounters above, treating 
their cumulative effect as a diffusion in phase space, this argument
would also apply to Monte-Carlo Fokker-Planck methods. However,
in a novel feature included recently by Fregeau, Rasio and 
co-authors \citep{fr2003,fr2007} in their Monte-Carlo method, 
both of the above types of encounters are handled in the following 
way. 

The dynamical evolution of the cluster is treated by a basically  
H\'enon-type Monte-Carlo method, which describes this evolution
as a sequence of equilibrium models, subject to regular velocity
perturbations which are calculated by the standard H\'enon 
method for representing the average effect of  many weak, 
frequent, distant encounters (see \citet{fr2003} and references
therein). In addition, the strong, rare, close encounters are 
by handled by (a) keeping track of the (Monte-Carlo-realized)
positions of the objects in the cluster, and so deciding whether
two given objects will undergo a strong, close encounter or not,
by a \emph{rejection method} very similar to that described 
above in Sec.~\ref{var}, and then (b) treating these encounters
first (i) through cross-sections compiled from analytic fits to 
numerical scattering experiments \citep{fr2003}, exactly as
we have done throughout our approach, and then, (ii) in a 
more detailed approach, through a direct integration of the 
strong interaction at hand using standard two- and three-body
integrators \citep{fr2007}. 

A direct comparison of our results
with those of above authors is, for the most part, not possible,
since we focused primarily on the formation, destruction and 
hardening of a compact binary population in a given GC 
environment, while Fregeau et. al focused primarily on the 
dynamical evolution of the GC environment in the presence 
of a given \emph{primordial} binary population. However, there is one
feature on which we were able to roughly compare our results
with those obtained by these and earlier authors. This is the 
problem of hardening of primordial binaries in GCs, pioneering 
studies which were performed through direct Fokker-Planck 
integration by \citet{G.et.al91}, and through Monte-Carlo  
method by \citet{H.et.al92}, and again recently through the 
above Monte-Carlo method by \citet{fr2003}.  
In an early test run of our scheme, we studied this
problem by ``turning off'' the binary formation and destruction 
terms in our scheme, thereby studying only the hardening of the
primordial binary population through our Boltzmann approach.
The results we obtained for the progressive hardening of the          
binary $a$-distribution profile (from an initial profile which was
uniform in $\ln a$, as in all the above references, and in our 
work)  were, indeed, very similar to those given in the above 
references.

In a pioneering study, \citet{dr92,dr94}
explored the tidal-capture formation and subsequent evolution 
of compact binaries in GCs, concentrating on recycled, 
millisecond pulsars in the first part of the study \citep{dr92},
and on CVs in the second part \citep{dr94}. These authors 
followed the histories of many neutron stars against a given 
background representing a GC core (parameters corresponding 
to 47 Tuc and $\omega$ Cen were used as typical examples),    
employing Monte-Carlo methods to generate tidal-capture
events in this environment. They followed the subsequent 
orbital evolution of these binaries due to hardening by 
gravitational radiation and magnetic braking, until Roche
lobe contact occurred. In those cases where such contact
occurred through orbit shrinkage before the low-mass 
companion could reach the giant phase due to its nuclear 
evolution, these authors did not follow further evolution of
the binary, while they did so when the contact occurred due
to the evolutionary expansion of the companion. 

From the above considerations, 
Di Stefano and Rappaport estimated the expected 
number of recycled pulsars and CVs in GCs like 47 Tuc and
$\omega$ Cen, and also gave the orbital-period distribution of 
the above binaries at two points, \viz, (a) just after tidal capture
and orbit circularization, and (b) at Roche-lobe contact. 
However, their orbital-period distributions cannot be compared
directly with those given in this paper (or Paper I) for the 
following reason. In the Monte-Carlo method of these authors,
tidal capture occurs at different times for different binaries, as
does Roche-lobe contact. Thus, showing the orbital-period 
distribution at any of the above two points means, in effect, that
the period-distributions at different times are being mixed. By
contrast, we have (in this paper and in Paper I) studied the 
evolution of the orbital period-distribution in time, displaying
``snapshots'' of the whole distribution at various times, which
we called ``time slices'' above and in Paper I. In our display, for
example, at any given time, some binaries are in Roche-lobe 
contact and some are not. Indeed, it seems that the orbital 
period-distributions just after tidal capture, as given by 
\citet{dr92}, should be compared with corresponding
N-body results given in \citet{z297}, and indeed they appear
rather similar.  We have, of course, pointed out in Paper I,
and stress the point here again, that our orbital 
period-distributions are to be regarded at this stage as 
intermediate steps in our calculation --- rather than final results
to be compared with future data on orbital period-distributions
of X-ray binaries in GCs --- because stellar-evolutionary effects
on binary evolution have not been included yet in our scheme
(also see below). With this inclusion, the aim would be to
produce the GC-analogue of such orbital period-distributions
as have been computed by \citet{prp03} for LMXBs 
\emph{outside} GCs.   
                       
In addition to the above improvement,  we listed in Paper I 
various other improvements and extensions that are to be 
implemented in our scheme in future. For example, the 
compact-binary distribution function above can be looked upon 
as one obtained by integrating the full, multivariate distribution 
function which includes other variables, \eg, the binding energy 
of the binary in the gravitational potential of the GC --- the 
so-called \emph{external} binding energy (or, equivalently, the 
position of the binary within the GC potential well
\citep{H.et.al92}), over these other variables. It would be most
instructive to be able to follow the evolution in these additional
variables in a more elaborate future scheme.

Encouraged by the veracity of the continuous limit, as presented
in this paper, we plan to conclude our program of the first stage 
of exploration of our Boltzmann scheme by studying one more
problem in the same spirit of demonstration of feasibility as we
have followed here and in Paper I. This is the question of 
compact-binary evolution in the environment of an 
\emph{evolving} GC. Whereas, in keeping with the tradition of 
numerous previous studies, we have treated the GC environment
in Paper I and here as a fixed (\ie, unchanging in time) stellar 
background, in reality a GC is believed to undergo considerable 
evolution following the long, quasi-static, ``binary-burning''   
phase, passing through phases of deep core collapse, (possible)
gravothermal oscillations, and so on. In this concluding study,
we propose to demonstrate that, at the current level of 
approximation in our scheme, and in the continuous limit, it is 
possible to follow the evolution of compact-binary populations
of GCs through these phases of GC evolution, at the expense of 
only a modest amount of computing time.



\acknowledgments

It is a pleasure to thank H. M. Antia, D. Heggie, P. Hut, S. Portegies 
Zwart, and F. Verbunt for stimulating discussions, and the referee  
for helpful suggestions.

\appendix

\section{Wiener Processes}

\emph{Wiener process} is the formal mathematical description of 
Brownian motion --- a classic example of a stochastic process, 
wherein a particle (\eg, pollen grain) on the surface of water 
undergoes random motion due to stochastic bombardment of it by 
water molecules. A standard description of the motion such a 
particle is given by the following differential form due to 
Langevin:
\begin{equation}
dX_t=a(t,X_t)dt+\sigma(t,X_t)\zeta_tdt.
\label{eq:brown}
\end{equation}
Here, $X_t$ is one of the components of the particle velocity at 
time $t$, and $a(t,X_t)$ is the retarding viscous force. The second 
term on the right-hand side represents the random molecular force, 
represented as a product of an intensity factor $\sigma(t,X_t)$ 
and a random noise factor $\zeta_t$, the latter at each time $t$ 
being a random number, suitably generated.

A standard Wiener process $W(t)$ is often defined as a continuous
Gaussian process with independent increments, satisfying the    
following properties:
\begin{equation}
W(0)=0,\qquad E(W(t))=0,\qquad {\rm Var}(W(t)-W(s))=t-s, 
\end{equation}
for all $0\le s\le t$. Here, $E$ represents the expectation value 
and `Var' the variance of the indicated stochastic 
variable\footnote{Strictly speaking, the first equation should be
written as $W(0)=0$, w.p.1, where `w.p.1' stands for 'with 
probability one', since we are dealing with random variables here.
But we shall not go into mathematical rigor here, referring the
reader to \citet{kl94}}. Note that a Wiener process $W_t(\omega)$, 
can also be thought of as a ``pure'' Brownian motion with $a=0$
in Eq.~(\ref{eq:brown}) \citep{kl94}, wherein the increments 
$dW_t(\omega)$ for any sample path $\omega$ represent a Gaussian 
white noise.  

Eqn.~(\ref{eq:brown}) can then be rewritten in terms of the 
\emph{symbolic differential} (see below) $dW_s(\omega)\equiv 
\zeta_s(\omega)ds$ of a Wiener process, and its integral form 
\begin{equation}
X_t(\omega)=X_{t_0}(\omega)+\int_{t_0}^{t}a(s,X_s(\omega))ds
+\int_{t_o}^t\sigma(s,X_s(\omega))dW_s(\omega)
\label{eq:brown2}
\end{equation}  
represents a path integral over the trajectory of the particle for 
the sample path $X_t(\omega)$, where $\omega$ is a particular 
trajectory of the Brownian particle.

\section{It\^{o} calculus} 

The problem with the second term on the right-hand side of 
Eqn.~(\ref{eq:brown2}), which represents an integral along a 
\emph{Wiener path}, is that it is not defined in ordinary calculus, 
since $W_t(\omega)$ is not differentiable in the ordinary sense.
Such an integral along a Wiener path has to be \emph{redefined}
suitably to become acceptable mathematically, and the It\^{o} 
integral is a well-known example of this. The classical limit-of-sum
definition of an integral does not hold good for an It\^{o} integral
like
\begin{equation}
X_t(\omega)=\int_{t_0}^t f(s,\omega)dW_s(\omega),
\label{eq:ito}
\end{equation}    
since the corresponding finite sum will be divergent over a Wiener 
path, as sample paths of a Wiener process do not have bounded 
variance (see above). However, it can be shown that such a sum
is \emph{mean-square convergent} under very general conditions 
\citep{ok2004}, owing to the well-behaved mean-square properties 
of Wiener processes. Accordingly, Eqn.~(\ref{eq:ito}) is defined 
only in the sense of \emph{mean-square convergence}, with the 
result that the integral (\ref{eq:ito}) is a random variable 
$X_t(\omega)$ with the following properties:
\begin{equation}
E(X_t)=0, \qquad E(X_t^2)=\int_{t_0}^{t}E(f(s)^2)ds
\end{equation}  

Consider now the well-known \emph{It\^{o} formula} for the 
transformation of a function $f(X_t)$ of a stochastic variable $X_t$
\citep{g95}. For simplicity, first assume that $X_t$ follows a 
stochastic equation of the form
\begin{equation}
X_t=X_{t_0}+\int_{t_0}^ta(X_t)dt + \int_{t_0}^t\sigma(X_t)dW_t,
\label{eq:brown3}
\end{equation}
\ie, the same as Eqn.~(\ref{eq:brown2}), but without explicit time 
dependence in the continuous and stochastic terms. For brevity,
we drop the symbol $\omega$, representing the sample path, in 
Eqn.~(\ref{eq:brown3}) and from here on. 
Let us divide the entire time span into time-steps at $t_1, t_2,
\ldots t_k,\ldots$ of length $h_1,h_2,\ldots h_k,\ldots$ with the 
largest step size $h_{max}$. Then $X_t$ at times $t_k$ and 
$t_{k+1}$ are related by
\begin{equation}
X_{k+1}=X_{k}+\int_{t_k}^{t_{K+1}}a(X_t)dt + \int_{t_k}^{t_{K+1}}
\sigma(X_t)dW_t,
\label{eq:brown4}
\end{equation}
where we write $X_k\equiv X_{t_k}$ and $X_{k+1}\equiv X_{t_{k+1}}$
for brevity. The It\^{o} formula states \citep{ok2004} that:
\begin{equation}
f(X_t)=f(X_k)+\int_{t_k}^t{\mathcal L}f(X_s)ds+\int_{t_k}^t 
f^\prime(X_s)\sigma(X_s)dW_s,
\label{eq:itorule}
\end{equation}
where the operator ${\mathcal L}$ is defined by:
\begin{equation}
{\mathcal L}f(X_s)\equiv f^\prime(X_s)a(X_s)+{1 \over 2}
f^{\prime\prime}(X_s)\sigma^2(X_s).
\label{eq:Ldef}
\end{equation}
For explicitly time-dependent continuous and stochastic terms, the
It\^{o} formula can be generalized suitably.
 
We can use Eqn.~(\ref{eq:itorule}) in Eqn.~(\ref{eq:brown4}) to expand 
$a(X_t)$ and $\sigma(X_t)$ around $t_k$:
\begin{equation}
\begin{array}{l}
X_{k+1}=X_k+a(X_k)h_{k+1}+\sigma(X_k)\Delta W_{k+1}\\
+\int_{t_k}^{t_{k+1}}\int_{t_k}^t{\mathcal L}a(X_s)dsdt + 
\int_{t_k}^{t_{k+1}}\int_{t_k}^ta^{\prime}(X_s)\sigma(X_s)dW_sdt\\
+\int_{t_k}^{t_{k+1}}\int_{t_k}^t{\mathcal L}\sigma(X_s)dsdW_s + 
\int_{t_k}^{t_{k+1}}\int_{t_k}^t\sigma^{\prime}(X_s)\sigma(X_s)dW_sdW_t
\end{array}
\label{eq:taylorexp}
\end{equation}
Now, if we discard all terms in Eqn.~(\ref{eq:taylorexp}) of 
${\mathcal O}(h^\alpha)$ for $\alpha>1$, we obtain
\begin{equation}
X_{k+1}=X_k+a(X_k)h_{k+1} + \sigma(X_k)\Delta W_{k+1} + 
{1\over 2}\sigma^{\prime}(X_k)\sigma(X_k)\left((\Delta W_{k+1})
^2-h_{k+1}\right),
\end{equation}
which is known as the \emph{Milshtein scheme}. This is the stochastic 
analogue of the second-order Taylor expansion of ordinary calculus. 
The Milshtein scheme can be shown to be \emph{strongly} or \emph{pathwise} 
convergent \citep{kl94} to order $h$, in the sense that the solution 
converges to the actual Brownian path as $h_{max}\rightarrow 0$.
If we restrict the expansion upto the ${\mathcal O}(h^{1/2})$ terms, 
\ie, upto the first three terms in the right-hand side of 
(\ref{eq:taylorexp}), we obtain a slower ($\sim h^{1/2}$) pathwise 
convergence, which is known as the \emph{Euler-Maruyama scheme}.

For higher dimensions, with $X_k\in{\mathcal R}^\Ncal$ and 
$W_t\in{\mathcal R}^\Dcal$, the second-order stochastic Taylor 
expansion of $X_k^i$ is given by (see \citet{g95} and references 
therein):
\begin{equation}
X_{k+1}^i=X_k^i+a^i(X_k)h_{k+1} + \sum_{j=1}^\Dcal\sigma_j^i(X_k)
\Delta W_{k+1}^j + \sum_{j=1}^{\Ncal}\sum_{p,q=1}^{\Dcal}
\frac{\partial \sigma_p^i}{\partial X^j}
\sigma_q^j(X_k)I_{pq}(k,k+1)+R,
\label{eq:taylor1}
\end{equation}
where
\begin{equation}
I_{pq}(k,k+1)\equiv\int_{t_k}^{t_{k+1}}\int_{t_k}^t dW_s^pdW_t^q
\end{equation}
and $R$ contains all terms of ${\mathcal O}(h^\alpha)$ for 
$\alpha>1$. If $d\leq p,q$ ($p\neq q$), we obtain upon integration
by parts:
\begin{equation}
I_{pq}(k,k+1)+I_{qp}(k,k+1)=\Delta W_{k+1}^p\Delta W_{k+1}^q 
\equiv B_{pq}(k,k+1).
\label{eq:bpq}
\end{equation}
If we further define
\begin{equation}
A_{pq}(k,k+1) \equiv I_{pq}(k,k+1)-I_{qp}(k,k+1),
\label{eq:apq}
\end{equation}
then we can, with the aid of Eqns.~(\ref{eq:apq}) and (\ref{eq:bpq}),
express $I_{pq}$ in terms of $A_{pq}$ and $B_{pq}$. Substituting the
result in Eqn.~(\ref{eq:taylor1}), we finally obtain,
\begin{align}
X_{k+1}^i= & X_k^i+a^i(X_k)h + \sum_{p}\sigma_p^i(X_k)\Delta W_{k+1}^p 
\nonumber\\
&+\frac{1}{2}\sum_{j=1}^{\Ncal}\sum_{p=1}^{\Dcal}\frac{\partial 
\sigma_p^i}{\partial X^j}\sigma_p^j(X_k)\left((\Delta W_{k+1}^p)^2 
-h_{k+1}\right) \nonumber\\
&+\sum_{j=1}^{\Ncal}\sum_{0<p<q\leq \Dcal}\frac{1}{2}\left
(\frac{\partial\sigma_q^i}{\partial X^j}\sigma_p^j+\frac{\partial
\sigma_p^i}{\partial X^j}\sigma_q^j\right)(X_k)B_{pq}(k,k+1)
\label{eq:taylor2}\\
&+\sum_{j=1}^{\Ncal}\sum_{0<p<q\leq \Dcal}\frac{1}{2}\left
(\frac{\partial\sigma_q^i}{\partial X^j}\sigma_p^j-\frac{\partial
\sigma_p^i}{\partial X^j}\sigma_q^j\right)(X_k)A_{pq}(k,k+1)+R
\nonumber
\end{align}
If $\forall{\rm~}i,p,q$ 
\begin{equation}
\sum_{j=1}^{\Ncal}\left(\frac{\partial\sigma_q^i}
{\partial X^j}\sigma_p^j-\frac{\partial\sigma_p^i}{\partial X^j}
\sigma_q^j\right)=0,
\label{eq:commute}
\end{equation}
then the $A_{pq}$ terms drop out of Eqn.~(\ref{eq:taylor2}). 
Equation~(\ref{eq:commute}) is called the \emph{commutativity 
condition} and is usually written as,
\begin{equation}
[\sigma_p,\sigma_q]=0.
\label{eq:comm}
\end{equation}
When the above commutativity condition is not satisfied, the 
quantities $A_{pq}$, known as the \emph{Levy areas}, have to be 
calculated in order to achieve second-order accuracy. 




\clearpage



\clearpage
%
%

\clearpage

\clearpage

\clearpage

\clearpage

\clearpage





\begin{thebibliography}{}


\bibitem[Banerjee \& Ghosh (2006)]{bg2006}
Banerjee, S. and Ghosh, P., 2006, \mnras, 373, 1188.

\bibitem[Banerjee \& Ghosh (2007)]{pI}
Banerjee, S. and Ghosh, P., 2007, \apj, 670, 1090 (Paper I).

\bibitem[Di Stefano \& Rappaport (1992)]{dr92}
Di Stefano, R. and Rappaport, S., 1992, \apj, 396, 587.

\bibitem[Di Stefano \& Rappaport (1994)]{dr94}
Di Stefano, R. and Rappaport, S., 1994, \apj, 423, 274.  

\bibitem[Fabian et.al. (1975)]{fpr75}
Fabian, A. C., Pringle, J. E. and Rees, M. J. 1975, \mnras, 172, 15P.

\bibitem[Fregeau et.al (2003)]{fr2003}
Fregeau, J. M., G\"urkan, M. A., Joshi, K. J. and Rasio, F. A., 2003,
\apj, 593, 772.

\bibitem[Fregeau \& Rasio (2007)]{fr2007}
Fregeau, J. M. and Rasio, F. A., 2007, \apj, 658, 1047.

\bibitem[Gains (1995)]{g95}
Gaines, J.G., 1995, in ``Stochastic Partial Differential Equations'',
ed. Etheridge, A., Cambridge University Press.  

\bibitem[Gao et.al (1991)]{G.et.al91}
Gao, B., Goodman, J., Cohn, H. and Murphy, B., 1991, \apj,
370, 567. 

\bibitem[Heggie (1975)]{h75}
Heggie, D., 1975, \mnras, 173, 729.

\bibitem[Heggie, Hut \& McMillan (1996)]{hhm96}
Heggie, D., Hut, P. and McMillan, S.L.W., 1996, \apj, 467, 359.

\bibitem[Hut \& Bahcall (1983)]{hb83}
Hut, P. and Bahcall, J.N., 1983, \apj, 268, 319.

\bibitem[Hut, McMillan \& Romani (1992)]{H.et.al92}
Hut, P., McMillan, S. and Romani, R.W., 1992, \apj, 389, 527.

\bibitem[Kloeden et.al (1994)]{kl94}
Kloeden, P.E., Platen, E. and Schurz, H., 1994, ``Numerical Solution of
SDE Through Computer Experiments'', Springer-Verlag.

\bibitem[Lee \& Ostriker (1986)]{lo86}
Lee, H.M. and Ostriker, J.P., 1986, \apj, 310, 176. 

\bibitem[Milshtein (1974)]{m74}
Milshtein, G. N., \emph{Th. Prob. Appl.}, 19, 557.

\bibitem[{\O}ksendal (2004)]{ok2004}
{\O}ksendal, B., 2004, ``Stochastic Differential Equations'',  
Springer International Edition. 

\bibitem[Pfahl et.al. (2003)]{prp03}
Pfahl, E.D., Rappaport, S. and Podsiadlowski, P. 2003, \apj, 597, 1036.

\bibitem[Pooley et.al (2003)]{pool2003}
Pooley, D. et.al., 2003, \apj, 591, L131.

\bibitem[Portegies Zwart et.al (1997a)]{z197}
Portegies Zwart, S. F., Hut, P. and Verbunt, F., 1997a, \aap, 328, 130.

\bibitem[Portegies Zwart et.al (1997b)]{z297}
Portegies Zwart, S.F., Hut, P., McMillan, S.L.W. and Verbunt, F., 1997b, 
\aap, 328, 143.

\bibitem[Press \& Teukolsky (1977)]{pt77}
Press, W.H. and Teukolsky, S.A., 1977, \apj, 213, 183.

\bibitem[Press et.al (1992)]{pr1992}
Press, W.H., Teukolsky, S.A., Vetterling, W.T., Flannery, B.P.,
1992, ``Numerical Recipes in C'', Cambridge University Press.

\bibitem[Shull (1979)]{sh79}
Shull, J.M., 1979, \apj, 231, 534.

\bibitem[Sigurdsson \& Phinney (1993)]{sp93}
Sigurdsson, S. and Phinney, E.S., 1993, \apj, 415, 631.

\bibitem[Spitzer (1987)]{spz}
Spitzer, L. Jr., 1987, ``Dynamical Evolution of Globular Clusters'', 
Princeton Univ. Press.

\bibitem[Verbunt (2003)]{v2002}
Verbunt, F., 2003, in ``New Horizons in Globular Cluster Astronomy'', 
ASP conf. series. 296, 245, eds. G. Piotto et al., 
Astron. Soc. Pacific, San Francisco.

\bibitem[Verbunt (2006)]{v2006}
Verbunt, F., 2006, in ``Highlights of Astronomy'', Volume 14, Proc.
XXVIth IAU General Assembly, Prague 2006, ed. van der Hucht, K.A.,
IAU Publ, Paris.

\bibitem[Verbunt \& Hut (1987)]{vh87}
Verbunt, F. and Hut, P., 1987, in IAU Symp. 125, ``The Origin and 
Evolution of Neutron Stars'', ed. D. J. Helfand \& J. H. Huang,
(Dordrecht: Reidel), 187.

\bibitem[Verbunt et.al (1989)]{v.et.al89}
Verbunt, F., Lewin, W.G.H and van Paradijs, J. 1989, 
\mnras, 241, 51.  



\end{thebibliography}
\end{document}